\documentclass[onecolumn]{aa} 
\usepackage{natbib}
\usepackage{graphicx}
\usepackage{txfonts}
\begin{document}
   \title{The ultraviolet properties of Luminous Infrared Galaxies at z$\sim$0.7}

   \subtitle{Is there any evolution in their dust attenuation?}

   \author{V. Buat
          \inst{1}
          \and
           D. Marcillac\inst{2}
\and
D. Burgarella\inst{1}\and E. Le Floc'h\inst{2,3}\thanks{SPITZER fellow, Institute for Astronomy, University of Hawaii}\and G. Rieke \inst{2}\and T. T. Takeuchi \inst{4}\and J. Iglesias-Par\'amo \inst{5}\and C. K. Xu\inst{6} 
          }

   \offprints{V. Buat}

   \institute{Observatoire Astronomique Marseille Provence, Laboratoire d'Astrophysique de Marseille,
BP8, 133761 Marseille cedex 12, France\\
              \email{veronique.buat@oamp.fr}
         \and
        Steward Observatory, University of Arizona, 933 North Cherry Avenue, Tucson, AZ 85721\\
           \and
Institute for Astronomy, University of Hawaii, 2680 Woodlawn Drive, Honolulu, HI 96822\\
          \and
Astronomical Institute, Tohoku University, Aoba, Aramaki, Aoba-ku, Sendai 
980--8578, Japan\\
\and
Instituto de Astrof\'{\i}sica de Andaluc\'{\i}a,CSIC, 18008 Granada, SPAIN\\
\and
California Institute of Technology, MC 405-47, 1200 East California 
Boulevard, Pasadena, CA 91125\\
 }

   \date{}

 
  \abstract
   {}
   {The  total infrared (TIR: 8-1000 $\mu$m) and far-ultraviolet (FUV: $\sim 1500\AA$) luminosity functions of galaxies and the related luminosity densities $\rho_{TIR}$ and $\rho_{FUV}$ are known to evolve with redshift from z=0 to z$\sim$1 but with a different rate:  the galaxy populations appear brighter in the past at both wavelengths but the evolution in TIR is larger than in FUV. It leads to an  increase of the ratio of TIR to FUV luminosity densities $\rho_{TIR}/\rho_{FUV}$ which can be interpreted as a global increase of the dust attenuation from z=0 to z$\sim$1. Our aim is to understand the origin of this increase: is it entirely due to  a variation  of the dust attenuation with the luminosity of the galaxies as seen as z=0 or are properties of galaxies evolving with the redshift?}
   {We focus on  infrared galaxies more luminous than $L_{\rm TIR}=10^{11} L\sun$ at z$\sim$0.7 observed by SPITZER/MIPS and we measure their ultraviolet emission at 2310 $\AA$ from  GALEX images.  These Luminous InfraRed Galaxies  (LIRGs) represent the bulk of the TIR luminosity density at intermediate redshift. The analysis of the ratio of TIR to FUV (rest-frame) luminosity ($L_{\rm TIR}/L_{\rm FUV}$)  enables us to discuss their dust attenuation and to compare it to what is found in the nearby universe for galaxies of similar infrared luminosity and selected in the same way}
   {Some evolution of  $L_{\rm TIR}/L_{F\rm UV}$ and therefore of  dust attenuation is found: LIRGs at z=0.7 span a larger range of $L_{\rm TIR}/L_{\rm FUV}$ ratios than at z=0 and their mean dust attenuation at FUV wavelengths is found $\sim$0.5 mag lower than for their local counterparts. The  decrease of  dust attenuation is found lower  than that  reported in other studies for   bright galaxies  selected in UV rest-frame at z=1 and 2. A semi-quantitative analysis is performed which accounts for the general increase of  dust attenuation with the bolometric luminosity of  galaxies: it is found that the slight  decrease of dust attenuation for LIRGs at z=0.7 remains consistent with the increase of $\rho_{TIR}/\rho_{FUV}$ with  redshift.}
   {}


   \maketitle
%

\section{Introduction}

Rest-frame far-ultraviolet and thermal infrared surveys are commonly used to probe  star formation in the universe and its evolution as a function of redshift z. The problem of  dust attenuation  of stellar light in galaxies is central in these analyses because  FUV emission is directly affected by it and    infrared emission  emission takes its origin in this process.  As a consequence the total infrared (TIR) to FUV flux ratio is a robust tracer of dust attenuation in star forming galaxies \citep[e.g.][]{buatxu,meurer, gordon}\\
 In a broad sense, both observations (FUV or TIR) have led to similar conclusions: a decrease of  star formation rate (SFR) from z=1 to z=0 \citep{flores, lefloch, schim} which is also found with other tracers of star formation \citep[e.g.][and references therein]{hopkins06}. Each wavelength range is sensitive to a specific galaxy population and recovering all  star formation from mono-wavelength observations (ultraviolet or infrared) appears to be difficult, even at low z. \cite{buat06} have shown that FUV and TIR surveys of the nearby universe do not exactly lead to the same sampling of galaxies. \\
This effect seems to be amplified at higher z.
Thanks to  recent surveys of SPITZER and  GALEX  
   the total infrared (TIR)  and  far-ultraviolet (FUV-1530$\AA$) luminosity functions are now available from z=0 to z=1 \citep{arn,lefloch}: a strong evolution of both luminosity functions is seen but the ratio of  luminosity densities  $\rho_{TIR}/\rho_{FUV}$ increases from z=0 to z=1 \citep{ttt2}. This increase might be explained, at least qualitatively: dust attenuation is found to increase   with the bolometric luminosity or SFR of  galaxies   in the nearby universe \citep{wang,buat98,sullivan,hopkins01,martin,buat05} but also at higher z \citep{reddy,lefloch,bell05}. Therefore the general brightening of the galaxies when z increases,  observed from the evolution of luminosity functions,  may also induce an increase of the global dust attenuation.\\

Nevertheless, studies of galaxy samples at  $z>0$ have led to rather controversial results about the amount of dust obscuration in bright distant galaxies. \citet{reddy} and  \citet{burgarella} performed a FUV (rest-frame) like selection at z=2 and 1 respectively and found a  dust attenuation   about ten times lower than in the nearby universe. Conversely,  samples selected in infrared by \citet{choi} or \citet{bell05} led to attenuations consistent with the relations found at z=0 between the TIR to FUV flux ratio and the total SFR  (or equivalentely the luminosity of young stars).
 From  GALEX and SWIRE observations of the ELAIS-N1 field \citet{xu06apjs}  found average trends  consistent with no evolution  between z=0 and z=0.6 of the TIR to FUV flux ratio of galaxies selected in FUV  or at 24 $\mu$m except a decrease by a factor $\sim$2 of the mean $L_{\rm TIR}/L_{F\rm UV}$ for infrared selected galaxies with $L_{\rm TIR}\sim 10^{11} L\sun$. However, the detection rate in FUV (resp. 24 $\mu$m) of galaxies selected at 24 $\mu$m (resp. FUV) was only 27 $\%$ (resp. 20$\%$) and the analysis of \citet{xu06apjs} almost entirely rely on stacking. 

Here we  analyse   ultraviolet properties  of Luminous InfraRed Galaxies  (LIRGs with $L_{\rm TIR} \ge 10^{11} L\sun$) at medium z. \citet{lefloch} have shown that LIRGs represent about half of the total infrared luminosity density at z$\sim$0.7 and $\sim 70\%$ at z$\sim$ 1 with only a minor contribution of Ultra Luminous InfraRed Galaxies (ULIRGs with $L_{\rm TIR} \ge 10^{12} L\sun$). On the contrary, most of UV selected galaxies   are expected to have a lower TIR luminosity  than LIRGs: at $z\sim 1$ \citet{burgarella}  found that only ~17$\%$ of their sample of  Lyman Break Galaxies are detected by SPITZER at 24 $\mu$m. A preliminary analysis of a FUV selection of galaxies at z=0.7 leads to only 10-15$\%$ of them with $L_{\rm TIR} \ge 10^{11} L\sun$) (Takeuchi et al., in preparation). However, even if they are not numerous, these UV selected galaxies also detected as LIRGs are found to be major contributors to the total star formation rate \citep{burgarella2} \\
We  select our sample from the deep SPITZER/MIPS survey of the Chandra Deep Field South \citep{lefloch} focusing on redshift range 0.6-0.8. GALEX has also surveyed this area and very deep images are available.  So we should have  a very high detection rate of  LIRGs with GALEX. 
In section 2, we  present the data selection and the measurement of the ultraviolet emission of  infrared sources. Then, in section 3,  we  analyse the ratio of the TIR to FUV luminosity ($L_{TIR}/L_{FUV}$) which is a tracer of  dust attenuation in galaxies. A comparison between z=0 and z=0.7 is  performed. In section 4 we compare our results to  other studies. Section 5 is devoted to  conclusions.
\\
Throughout this article, we use the cosmological  parameters $H_0 = 70$ km s$^{-1}$ Mpc$^{-1}$, 
$\Omega_M = 0.3$ and $\Omega_{\lambda} = 0.7$. All magnitudes will be quoted in the AB system.
The TIR luminosity $L_{\rm TIR}$ is defined on the wavelength range 8-1000 $\mu$m. The FUV luminosity $L_{\rm FUV}$ is defined as $\nu L_{\nu}$ with $L_{\nu}$ in $\rm erg ~cm^{-2} s^{-1} Hz^{-1}$

\section{The data}

\subsection{Selection of the sample at z $\sim 0.7$}

We start with the sample of galaxies in the Chandra Deep Field South (CDFS) used by \citet{lefloch} to build  TIR luminosity functions from z=0 to z=1. 
It consists of 2955 sources detected at 24 microns by MIPS with $f_{24}$$>$83$\mu$Jy at the 80 \% completeness limit (Papovich et al., 2004).
As explained in \citet{lefloch},  several spectroscopic surveys  \citep{lefevre, vanzella,szokoly} are primarily used to associate a spectroscopic redshift to the MIPS sources.
Photometric redshifts from COMBO-17 \citep{wolf} are also used for sources  at z$\leq$1.2 and brighter than $R_{Vega}$ $\sim$ 24, where they are accurate enough ($\delta_z$/(1+z)$\leq$10\%) to estimate  quantities. We refer to \citet{lefloch} for more details. 

 As discussed below, our strategy  consists in measuring the NUV (2310$\AA$) emission of these galaxies directly on the GALEX images at the position of the MIPS sources. We  take a special care to avoid contaminations  by   sources in the close vicinity of  MIPS ones which can also be ultraviolet emitters.  At this aim, we apply strict selection criteria to the initial sample. All the steps followed to build the final sample and described  in the sections 2 and 3 are summarized in Table ~\ref{table1}.
First, we work at z$\sim$0.7 and  only sources with a redshift comprised between 0.6 and 0.8 are kept; 623 galaxies are thus selected. A single optical counterpart must be found in COMBO-17 within 2 arcsec from the coordinates of the MIPS source. This tolerance radius was also adopted by \citet{lefloch} for their identifications. This  choice is motivated by the astrometrical precision of  MIPS/24 $\mu m$ images and the rather large FWHM of  MIPS 24 $\mu$m PSF ($\sim 6$ arcsec). It also accounts for a potential physical shift between the infrared and optical emission of disturbed objects. Here, we add an additional criterium: we  exclude all MIPS sources associated to two  or more optical sources within 2 arcsec. 65 sources are dropped and  558 sources are left.

We have not a complete sample of galaxies selected at 24 $\mu$m but we expect to have no strong bias in the selection. Our selection of "isolated" sources (i.e. with only one optical source within 2 arcsec) might avoid   mergers (2 arcsec correspond to 14 kpc at z=0.7). Anyway, we cannot  risk to have the FUV emission coming from another source than the MIPS one. Moreover, the reference sample at z=0 used for comparison is built in the same way (cf section 3.1).
In order to check the effects of the exclusion of confused objects we have performed all the following analysis including  confused sources both at high and low z and the results have been found  unchanged.
\begin{table}
\caption{Number of sources at $0.6<z<0.8$ for the different selections applied to the original sample of galaxies selected at 24 $\mu$m. Each row of the table corresponds to an additional selection applied to the original sample of 623 sources (from top to bottom). The final sample of LIRGs is described in the last row of the table }\label{table1}
\centering
\begin{tabular}{lcc}
\hline \hline
  & number of sources at 24 $\mu$m & NUV detections   \\
original sample& 623& $-$ \\
\hline
a single optical counterpart& 558 &$-$ \\
reliable NUV photometry& 402& 331\\
LIRGs&190&158\\
\hline
\end{tabular}
\end{table}
\subsection{ NUV measurements of the 24 $\mu m$ sources}

GALEX \citep{morissey} observed the CDFS  for 76 ks  in both the FUV (1530 $\AA$) and the NUV (2310 $\AA$) as part of its deep imaging survey. The GALEX field of view (diameter $\rm 1.25 ~deg$) is centered at $\rm \alpha = 03h 32m 30.7s, \delta = -27 deg 52' 16.9"$. The resolution of the NUV image is measured and the FWHM of the PSF is found equal to 4.5 arcsec. Prior to any measurement we have applied a median filter $3 \times 3$ on the GALEX images which enlarges the PSF up to $\sim 6$ arcsec and makes it similar to that of the  MIPS image.\\

We use DAOPHOT \citep{stetson} to measure the NUV emission at the location of the 24 $\mu m$ sources. DAOPHOT was also used by \citet{lefloch} to measure  24 $\mu$m fluxes.  DAOPHOT is  well suited for point sources (stellar fields). We must check that we can use it on our GALEX field. \citet{demello} have measured the size of UV selected galaxies at intermediate redshifts. In the  redshift range 0.6-0.8  $\sim 90\%$ of their observed galaxies  have an  effective (half light) optical radius $R_e$ lower than 0.8 arcsec  (which corresponds to $\sim$ 6 kpc at z=0.7). If we assume a Gaussian distribution for the galaxy light, a  galaxy with $R_e = 0.8$ arcsec convolved with the GALEX PSF (also assumed to be Gaussian) would appear with a FWHM $\sim$ 6.7 arcsec. The photometry is performed on the central 3 arcsec (i.e. 2 pixels for the GALEX images) which  encloses  44 $\%$ of the total energy for a pure PSF (6 arcsec after $3 \times 3$ median filtering, see above). For the largest objects ($R_e = 0.8$ arcsec) the aperture of 3 arcsec would enclose 40$\%$ of the total energy. Therefore we estimate the photometric  uncertainty assuming all the sources punctual to be at most 4 $\%$  \\

  The PSF was built using 10 stellar like objects with a FUV magnitude comprised between 17.5 and 19.8 mag.
The GALEX field is rather dense with an average of $\sim$ 0.06 galaxy per beam. The influence of close neighbours in the measure of  NUV emission has been tested by adding artificial sources (ADDSTAR task): since we have excluded all the original MIPS sources with more than one optical source within 2 arcsec we simulated NUV point sources in the GALEX images located from 2 to 4 arcsec from the objects to be measured. As long as the galaxy  has a magnitude lower or equal to NUV = 24 mag the contamination due to  neighbours remains low (less than 0.4 mag in the worst cases) but when  the   source  is fainter than  NUV= 25 mag the contamination by  neighbours can be very high (reaching 1  mag or more), whatever the magnitude and the location of the contaminating source is. Therefore we decided to exclude the sources whose measured NUV emission is fainter than NUV = 24.5 mag and which have at least one  optical neighbour within 4 arcsec from the MIPS coordinates. At the end we are left with 402 sources for which the measure of  NUV flux is considered as reliable. 156 galaxies  are excluded because they are fainter than  NUV= 24.5 mag and with an optical neighbour within 4 arcsec. As for the confused sources we have checked that including the sources with a non reliable  NUV photometry (according to our strict criteria) does not modify the results of the subsequent analysis.

331 out of the 402 galaxies are detected in NUV. When the MIPS source is not detected by GALEX we put an upper limit on the  NUV mag: NUV=26.2 mag. It corresponds to a completeness of 80$\%$ and a photometric error of 0.07 $\pm 0.04$ mag. This limit is obtained by simulations of 500 artificial sources added to the original GALEX image (ADDSTAR task). The NUV magnitudes are corrected for foreground Galactic extinction using the dust map of \citet{schlegel} and the Galactic extinction curve from \citet{cardelli}.

It might be worth noting that only $\sim 20\%$ of the NUV sources at z$\sim$0.7 have a MIPS counterpart brighter than 83 $\mu$Jy. However, even if these galaxies are far from being   a dominant population (in number) for a UV selection, their contribution to the  total star formation rate of a UV selected  sample  is  likely to be important because of their very high luminosity:  \citet{burgarella2}  show that they account for $\sim$2/3 of the total star formation rate of  a sample of Lyman Break selected  at z$\sim$1. A full analysis of a FUV (rest-frame) galaxy sample at z=0.7 is underway (Takeuchi et al., in preparation).

\section{ The $L_{TIR}/L_{FUV}$ ratio of Luminous InfraRed Galaxies}

We focus on Luminous InfraRed Galaxies (LIRGs, brighter than $10^{11} L\sun$) because the SPITZER/MIPS sample is complete up to  z=0.8 for $L_{TIR}>10^{11} L\sun$ and these LIRGs are found to be at the origin of the bulk of the TIR emission at medium z \citep{lefloch}. 
Our aim is to compare  the properties of a sample of LIRGs at z=0.7 to  those of a reference sample taken at z=0 which will be defined in section 3.1.
Indeed, from z=0 to z=0.7
$\rho_{TIR}$ and $\rho_{FUV}$ are found to increase by a factor equal to $\sim$8 and $\sim$4 respectively, leading to a net increase of $\rho_{TIR}/\rho_{FUV}$ by a factor $\sim$2 \citep{ttt2}. So, we can expect to find some evolution (if any) on the properties of  galaxies in this redshift range.\\
We  start by the description of the reference sample in order to define the quantities to be compared with the z=0.7 sample. 
\subsection{ The reference sample at z=0} 

\citet{buat06} have built a sample of galaxies from the IRAS PSCz cross-correlated with the GALEX All Sky Imaging Survey (AIS)  over more than 2000 deg$^2$. It is a flux limited sample (f$_{60}>$0.6 Jy) of $\sim 700$ galaxies, most of them have a measured FUV flux at 1530 $\AA$ from the GALEX All sky Imaging Survey.  

The total infrared (TIR) emission of the galaxies was estimated from their emission at 60 and 100 $\mu$m (see \citet{buat06} for more details). From this sample  we select only LIRGs ($L_{\rm TIR}>10^{11} {\rm L\sun}$). We must   restrict the sampled volume  to be sure to detect all LIRGs within this volume. The limit $L_{\rm TIR}=10^{11} L\sun$ corresponds to $L_{60}= 0.47~10^{11} L\sun$ for a mean $L_{TIR}/L_{60}$ ratio of 2.13 (mean value found for the LIRGs of  our reference sample). With a flux limit at 60 $\mu$m in the PSCz of 0.6 Jy, we must truncate the sample to $\rm v<~16 000 ~{\rm km s}^{-1}$. 98 LIRGs are selected, 91 have a measured FUV flux, 7 are not detected by GALEX in FUV and an upper limit at FUV = 20.5 mag is adopted corresponding to a 3$\sigma$ detection limit for the GALEX-AIS survey \citep{morissey}.The criterium applied at z=0 to avoid confused sources  was  the absence of any neighbour within 1 arcmin from the IRAS source \citep{buat06}: at the distance of the selected LIRGs it corresponds to a projected distance larger than 14 kpc. Therefore we can assume that a similar criterium is applied at z=0 and z=0.7 (c.f. section 2.1) to select isolated sources.

\subsection{ The z=0.7 sample of LIRGS}

We choose to work at z=$\sim$0.7 because  there is an  over density of galaxies at this redshift in the Chandra Deep Field South \citep[e.g.][]{wolf}.    190 galaxies of our sample of 402 galaxies are selected as Luminous InfraRed Galaxies with $L_{\rm TIR} > 10^{11} L\sun$, the mean redshift of the LIRGs sample is $<z>=0.70\pm 0.05$. 158 out of  the 190 LIRGs (i.e. 83$\%$) are detected in NUV. 

The total infrared (TIR) emission  of these galaxies has been calculated by \citet{lefloch} from their emission at 24 $\mu$m and  we adopt the results of their calculations.
At z=0.7 the GALEX NUV band (2310 $\AA$) corresponds approximately to the GALEX FUV one (1530 $\AA$) adopted for the reference sample (1358 $\AA$ for the rest-frame UV emission observed in NUV at z=0.7 against 1530 $\AA$  for the GALEX FUV band). 
The amplitude of K-corrections are investigated. The UV continuum is assumed to be well described by a power-law $f_{\lambda}\propto \lambda^{~\beta}$\citep[e.g.][]{calzetti},   where $f_{\lambda}$ is expressed in $\rm erg ~cm^{-2} ~s^{-1} \AA^{-1}$ (or $f_{\nu}\propto \nu^{-\beta-2}$ where $f_{\nu}$ is expressed in $\rm erg ~cm^{-2} ~s^{-1}~ Hz^{-1}$). $\beta=-1$ corresponds to a flat distribution in $\nu\times f_{\nu}$ (no K-correction). Unfortunately $\beta$ cannot be measured for our sample at z=0.7. At z=0, $\beta$ can be deduced from the FUV-NUV color measured by GALEX \citep[e.g.][]{seibert}.   $\beta$ is found to vary from -1.5 to 1 which induces  $-0.03\le~\log(\nu~ f_{\nu})_{\rm FUV rest-frame}-\log(\nu~ f_{\nu})_{\rm NUV observed}~\le 0.10$. Therefore we prefer not to apply any K-correction to the data: we consider the NUV fluxes observed at z=0.7 as rest-frame FUV fluxes  directly comparable with the FUV fluxes observed at z=0.\\

\subsection {$L_{TIR}/L_{FUV}$ distributions}

 $L_{TIR}/L_{FUV}$ is a robust tracer of dust attenuation in star forming galaxies. Quantitative estimates can be made regardless of  details of the geometry and star formation history as long as  galaxies are still forming stars actively \citep[e.g.][]{buatxu,meurer,gordon,calzetti}. \citet{buat05} have performed calibrations for the GALEX bands. We reproduce their formulae for the FUV band:
\begin{eqnarray}
 A\mbox{(FUV) [mag]} &=& -0.0333~\left(\log \frac{L_{\rm TIR}}{L_{\rm FUV}}
   \right)^3 
   +0.3522~\left(\log\frac{L_{\rm TIR}}{L_{\rm FUV}}\right)^2
   +1.1960~\left(\log\frac{L_{\rm TIR}}{L_{\rm FUV}}\right)+0.4967 \;
\end{eqnarray}

\begin{figure}
   \centering
   \includegraphics[angle=-90,width=15cm]{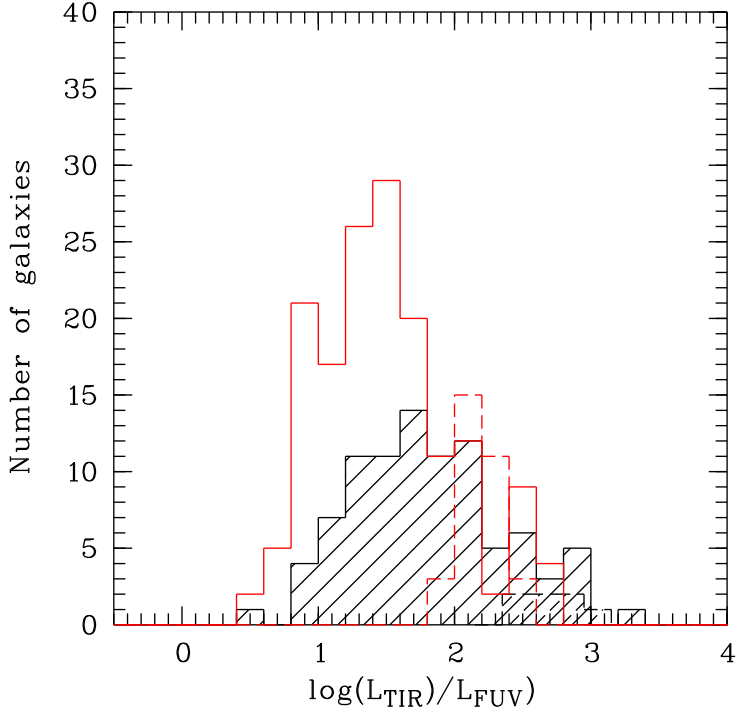}
      \caption{$L_{TIR}/L_{FUV}$ distribution for LIRGs at z=0 (black, hatched histogramm) and z=0.7 (red, empty histogramm), for both samples  histogramms of upper limits are plotted with  dashed lines}
              
         \label{hist}
   \end{figure}

$L_{\rm TIR}/L_{\rm FUV}$ has been found to increase with the bolometric luminosity of young stars in galaxies from z=0 to at least z=2 \citep [e.g.][and references in the introduction]{buat06}. As a matter of consequence, a comparison of  $L_{\rm TIR}/L_{\rm FUV}$ distributions can only be valid for galaxies exhibiting the same luminosity distribution. Our selection of LIRGs at z=0 and at z=0.7 compensates for the evolution of the TIR luminosity function reported by \citet{lefloch} which lead to brighter galaxies at higher z. Indeed the distributions of $L_{\rm TIR}$ in both samples are found very similar with $<log(L_{\rm TIR}/L\sun)>=11.22 \pm 0.17$ at z=0 and $<log(L_{\rm TIR}/L\sun)>=11.24 \pm 0.20$ at z=0.7. Therefore we can safely compare  $L_{\rm TIR}/L_{\rm FUV}$ distributions from both samples.\\
The histogramms of $L_{\rm TIR}/L_{\rm FUV}$ at z=0 and z=0.7 are presented in Fig.~\ref{hist}. We perform statistical tests accounting for non detections to compare the distributions (IRAF/stsdas/analysis/statistics/twosampt task). The two distributions are found to be drawn from different parent populations  with a probability larger than 0.99. The Kaplan Meier estimates of their mean are  slightly different: $<(log(L_{\rm TIR}/L_{\rm FUV})_{z=0.7}>=1.673 \pm 0.044$ and $<log(L_{\rm TIR}/L_{\rm FUV})_{z=0}>=1.897 \pm 0.063 $. Note that the uncertainty on  K-corrections (cf. section 3.2) does not question the robustness of the result since it  implies in most cases an under-estimate of $L_{\rm FUV}$ (at most 0.1 dex) and therefore an over-estimate of $L_{\rm TIR}/L_{\rm FUV}$ at z=0.7.\\
If we translate these mean values in quantitative measurements of  dust attenuation at FUV wavelength with formula (1)
we find $<A({\rm FUV})>=3.33 \pm 0.08 ~{\rm mag}$ at z=0.7 and $<A(\rm FUV)>=3.81 \pm 0.13 ~{\rm mag}$ at z=0. So a slight difference in the mean dust obscuration can be inferred from these data and the shapes of the distributions are significantly different:  the   distribution of $L_{\rm TIR}/L_{\rm FUV}$ at z=0.7 appears broader than at z=0 with an extension toward low $L_{\rm TIR}/L_{\rm FUV}$ values. 

The difference found between the mean $L_{\rm TIR}/L_{\rm FUV}$ values is slight. Therefore one may question the presence of some systematic effects in the derivation of the total infrared luminosity from the observables. Indeed, at z=0.7 $L_{\rm TIR}$ is deduced from the flux observed at 24 $\mu$m (14 $\mu$m in the rest-frame at z=0.7) by \citet{lefloch} using a set of local templates whereas at z=0 $L_{\rm TIR}$ is calculated with a combination of IRAS fluxes at 60 and 100 $\mu$m \citep{buat06}. 
\citet{ttt1} have proposed a formula to derive $L_{\rm TIR}$  from the luminosity at 15 $\mu$m (rest-frame) fully consistent with the IRAS database. To check the robustness of our present results we used the calibration of \citet{ttt1} to derive $L_{\rm TIR}$ at z=0.7 from the observed fluxes at 24 $\mu$m: we found the results unchanged with $<(L_{\rm TIR}/L_{\rm FUV})_{z=0.7}>=1.660 \pm 0.042$.
One must remind, however, that our analysis is based on the hypothesis that local galaxy templates are valid to estimate the total infrared luminosity of high redshift systems \citep{lefloch,marcillac}.
\section{Discussion}

\subsection{Comparison with previous works}
 
\begin{figure*}
   \centering
\includegraphics[angle=-90,width=18cm]{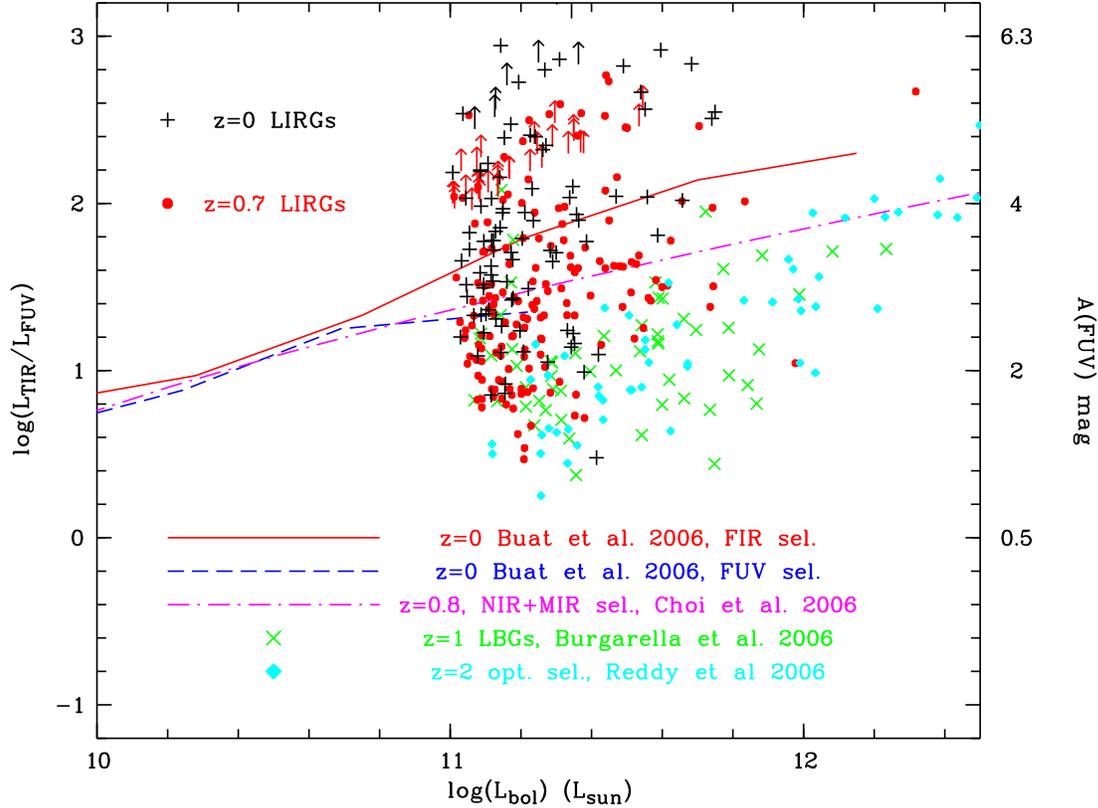}
      \caption{$L_{\rm TIR}/L_{\rm FUV}$ versus $L_{\rm bol} = L_{\rm TIR}+L_{\rm FUV}$. Vertical right axis: the dust attenuation $A({\rm FUV})$ is calculated with formula (1) in the text. LIRGs are plotted with  black 'plus' and arrows for our z=0 sample and  red dots and arrows for our z=0.7 sample. Blue dashed line: FUV selected sample at z=0 from \citet{buat06}, red solid line: FIR selected sample  at z=0 from \citet{buat06}. Dot dashed magenta line from \citet{choi}. The data from   \citet{reddy} are plotted with cyan lozanges, and those of\citet{burgarella,burgarella2} with green  crosses}
\label{ltir_lfuvfit}
   \end{figure*}

In Fig ~\ref{ltir_lfuvfit} is plotted the bolometric luminosity $L_{\rm bol}$, defined  as $L_{\rm TIR}+L_{\rm FUV}$,  versus $L_{\rm TIR}/L_{\rm FUV}$ for both samples at z=0 and z=0.7. The difference in the distributions is clearly seen with an extension to moderate $L_{\rm TIR}/L_{\rm FUV}$ values at z=0.7 not present at z=0.\\
For comparison, we have also gathered the results of previous works on  the  variation of $L_{TIR}/L_{FUV}$  at low and high z. In the nearby universe, we refer to \citet{buat06} and report here the mean relations found in their paper which are  consistent with other studies of local galaxies \citep[see][for more discussions]{buat06}. \\
At higher z   several recent studies are available. 
\citet{choi} have selected galaxies at z$\sim$0.8 at NIR+MIR wavelengths. By comparing the SFR deduced from the TIR emission and the strength of the emission lines they measured the extinction in the optical emission lines and found that the corresponding visual extinction varies as $A_{V} = 0.75 \log(L_{\rm TIR}/L\sun)-6.35 ~{\rm mag}$. To compare these results to the present work,
we have to translate this visual extinction in the gaseous medium to an attenuation for the ultraviolet stellar continuum. To be consistent with  \citet{choi}, we follow the Calzetti et al. recipe \citep{calzetti,calzetti01}. 
For the emission lines we adopt a foreground like distribution with a Milky Way extinction curve for a diffuse medium \citep{cardelli}: 
$$ A_{\rm V}=  3.1 ~ E(B-V)_{\rm g}$$  where $E(B-V)_{\rm g}$ is  the color excess for the  gas emission lines. 

  The color excess for the stellar continuum $E(B-V)_{\rm s}$ is given by $$E(B-V)_{\rm s} = 0.44 ~E(B-V)_{\rm g}$$ \citep{calzetti, calzetti01}. Then using the reddening curve \citep{calzetti}, it comes
$$ A_{\rm FUV} = k'(1530\AA)~ E(B-V)_{\rm s} = 10.33\times 0.44\times (A_{\rm V}/3.1)$$ with $k'(1530\AA) = 10.33$ which gives $ A_{\rm FUV} = 1.47 A_{\rm V}$.\\
  $A_{\rm FUV}$ is linked to $L_{\rm TIR}/L_{\rm FUV}$ via formula (1). 
The result   of these transformations is shown in Fig. \ref{ltir_lfuvfit}, note that in the Choi et al. relation the x axis is $L_{\rm TIR}$ and not $L_{\rm TIR}+L_{\rm FUV}$: for the LIRG regime (the topic of the present work) the difference is very small. The relation of \citet{choi} is consistent with the z=0 results at intermediate luminosity. For LIRG luminosities and higher the relation  flattens and is below the  relation found for FIR selected galaxies at z=0 (and consistent with the FUV selection at z=0). This trend is fully consistent with the present work and  the presence of LIRGs exhibiting a lower $L_{\rm TIR}/L_{\rm FUV}$ ratio.\\
Therefore, a difference is found between the amount of dust attenuation in LIRGs selected in infrared surveys at z=0 and z=0.7. The mean value of $L_{\rm TIR}/L_{\rm FUV}$ varies by 0.2 dex which translates in a decrease of  dust attenuation of $\sim 0.5$ mag from z=0 to z=0.7. The distribution of  $L_{\rm TIR}/L_{\rm FUV}$ is broader at z=0.7 than at z=0: a population of LIRGs with a moderate $L_{\rm TIR}/L_{\rm FUV}$ ($\log(L_{\rm TIR}/L_{\rm FUV}) \sim 1-1.5$ corresponding to an attenuation of $\sim 2-3$ mag in FUV) appears at z=0.7, this population is not  important at z=0. \\
A decrease of the dust attenuation can be linked to a lower metallicity in high redshift systems. Indeed, \citet{liang} found a mean metallicity of LIRGs at z$>$0.4 that is 0.3 dex lower as compared to that of local bright disks.
The attenuation of the UV stellar continuum was found to be correlated with metallicity in starburst and normal  galaxies  at z=0 \citep{cortese, heckman}. Using the relation given by \citet{cortese}: $\log(L_{\rm TIR}/L_{\rm FUV}) = 1.37 (12+\log(O/H)-11.36$ we obtain a decrease of $L_{\rm TIR}/L_{\rm FUV}$ of 0.4 dex from z=0 to z=0.7 consistent with what is found in Fig . \ref{ltir_lfuvfit}.\\
 The observed decrease of  dust attenuation in some LIRGS from z=0 to z$\sim$0.7 might also be related to the evolution of  morphological type for this galaxy population. From z=0 to z$>0.5$ \citet{melbourne} found a decrease of the number of peculiar/irregular systems exhibiting tidal features, assymetry or being obvious mergers as compared  to spirals,  this result is   confirmed by \citet{wang06} and \citet{bell05}. If disturbed galaxies are related to merging systems a larger dust attenuation is expected for them \citep[e.g.][]{sanders} and their  lower contribution  to LIRGs at high z as compared to low z might imply a decrease of the mean dust attenuation  for these galaxies.\\

At z=2 \citet{reddy}  studied  star formation and dust obscuration of galaxies predominantly selected in optical (UV rest-frame). We select  galaxies with $L_{\rm TIR}~>~10^{11} L\sun$ and with $1.9<z<2.3$ from their sample \citep{reddy2}. In this redshift range the G-band corresponds to the FUV-band of GALEX in the galaxy rest-frame. We plot these selected data in Fig . \ref{ltir_lfuvfit}: they appear to be distributed below our infrared selection. A comparison of the $L_{\rm TIR}/L_{\rm FUV}$ distributions is made difficult because of the obvious difference in the luminosity distributions of our samples at z=0 or 0.7 and the Reddy et al.  sample at z=2 which contains a relatively large number of ULIRGs ($L_{\rm TIR}~>~10^{12} L\sun$). 
If  the comparison is restricted to galaxies with $10^{11}<L_{\rm TIR}~<~10^{12} L\sun$, the mean $log(L_{\rm TIR}/L_{\rm FUV})$ found at z=2 is 0.9 corresponding to  1.85 mag, i.e.  $\sim$1.5 magnitudes lower than our  mean value at z=0.7 for our selection of LIRGs at 24 $\mu$m. The discrepancy seems to be lower for the brightest galaxies (ULIRGs). Given the low number of such bright galaxies in our sample at z=0.7 we cannot make any quantitative comparison; nevertheless in their study of the ELAIS-N1 field \citet{xu06apjs} also found that  dust obscuration in ULIRGs at z=0.6 is consistent with that obtained at z=0. \\
At z=1,  \citet{burgarella} studied Lyman Break Galaxies (i.e. GALEX FUV dropouts) and also found  a low obscuration for the galaxies they detected at 24 $\mu$m (only $\sim 15 \%$ of their sample is detected at 24 $\mu$m). In Fig. \ref{ltir_lfuvfit} we report their data \citep{burgarella2}. The obscuration that they obtain is consistent with that found by \citet{reddy} at z=2.\\
How to reconcile the results of \citet{reddy} and \citet{burgarella} at z$\sim$1 \& z$\sim$2 with ours at z=0.7? 
First we can invoke an evolution with the redshift but z=1 and z=0.7 are separated by only 1.5 Gyr and it is difficult to expect a large evolution during such a short timescale. 
The most natural explanation is to invoke  selection effects since we have a selection at 24 $\mu$m (15 $\mu$m rest-frame) whereas the selections of \citet{reddy} and \citet{burgarella} are predominantly in the rest-frame ultraviolet. Nevertheless at z=0 \citet{buat06} find only a slight difference between a TIR and a FUV selection so we 
  must assume a strong evolution of the properties of intrinsically luminous galaxies with z.\\

\subsection{Do we expect a decrease of dust attenuation for LIRGs at $z \sim 0.7$?}

At first glance the answer to this question is no: we do not expect a decrease of dust attenuation for bright galaxies at $z \sim 0.7$  since
(as underlined in the introduction) the ratio of the luminosity densities  
$\rho_{\rm TIR}/\rho_{\rm FUV}$ increases with $z$. 
Nevertheless, we observe a slight decrease of dust attenuation at a fixed $L_{\rm bol}$ and one must  also account with the intrinsic brightening of the galaxies when z increases together with  
the increase of dust attenuation with the bolometric luminosity of  galaxies observed at low and high z (cf. Fig. \ref{ltir_lfuvfit}). Are all these trends consistent with each other?

 We consider these issues according to a rather crude and semi-quantitative analysis  
\citep[e.g.][]{xu06apjs}.
On one hand the evolution of luminosity densities in TIR and FUV from $z=0$ to 
$z=0.7$ has been quantified: \citet{lefloch} found that $\rho_{\rm TIR}$ 
increases as $(1+z)^{3.9\pm 0.4}$ and \citet{schim} obtained 
$(1+z)^{2.5\pm 0.7}$ for the evolution of $\rho_{FUV}$. 
This gives an evolution of 
$\rho_{\rm TIR}/\rho_{\rm FUV}\propto (1+z)^{1.4\pm 0.8}$. 
Therefore from $z=0$ to $z=0.7$ an increase of 
$\rho_{\rm TIR}/\rho_{\rm FUV}$ by a factor $2.1_{-0.7}^{+1.1}$ is 
obtained.

On the other hand, we can predict very crudely the evolution of dust 
attenuation of a typical $L_{\rm TIR}^\star$ galaxy from the relation 
between dust attenuation and the bolometric luminosity of galaxies obtained at $z=0$. 
This is regarded to represent the global trend of the evolution of  
galaxies selected in infrared, since \citet{lefloch} showed that the evolution of TIR
luminosity function is approximately described by the pure luminosity 
evolution (PLE) with a small density evolution.
For the purpose of the discussion, let us perform a linear regression between $L_{\rm TIR}/L_{\rm FUV}$ and $L_{\rm bol}$  at $z=0$: at this aim we take the mean values used to plot the solid and dashed lines at z=0  in Fig. \ref{ltir_lfuvfit} (see \citet{buat06} for more details) and  
we find 
\begin{eqnarray}
  \log \left( \frac{L_{\rm TIR}}{L_{\rm FUV}}\right) = 
  0.64~\log (L_{\rm bol}/L\sun) - 5.5 \;.
\end{eqnarray}
If we assume that $L_{\rm bol}$ is not very different of $L_{\rm TIR}$ 
\citep{buat06}, it translates to 
\begin{eqnarray}
  \frac{L_{\rm TIR}}{L_{\rm FUV}} \propto L_{\rm TIR}^{0.6} \;.
\end{eqnarray}
As mentioned above, \citet{lefloch} found that the typical TIR luminosity 
$L_{\rm TIR}^\star$ (the knee of the luminosity function) evolves as 
$(1+z)^{3.2^{+0.7}_{-0.2}}$, according roughly to the PLE.
This gives 
\begin{eqnarray}
  \frac{L_{\rm TIR}}{L_{\rm FUV}} \propto (1+z)^{1.9^{+0.4}_{-0.1}} \;.
\end{eqnarray}
Hence, if the $L_{\rm TIR}/L_{\rm FUV}$-$L_{\rm bol}$ relation found at $z=0$ 
is still valid at $z=0.7$, a 'typical' TIR selected galaxy brightens with $z$ 
and would have its $L_{\rm TIR}/L_{\rm FUV}$ increasing by a factor of 
$\sim 3$ ($2.6\mbox{--}3.4$).
This factor is, however, slightly higher than that found for the evolution of 
$\rho_{\rm TIR}/\rho_{\rm FUV}$ ($1.4\mbox{--}3.2)$.

These crude estimations show that a slight decrease of the mean 
$L_{\rm TIR}/L_{\rm FUV}$ from $z=0$ to $z=0.7$ is not inconsistent with 
the evolution of $\rho_{\rm TIR}/\rho_{\rm FUV}$ in the same redshift range. 
Moreover \citet{xu06apjs} used a steeper regression between 
$L_{\rm TIR}/L_{\rm FUV}$ and  $L_{\rm bol}$ (or equivalently SFR) 
and concluded to a larger discrepancy, the local relation between 
$L_{\rm TIR}/L_{\rm FUV}$ and $L_{\rm bol}$ predicting too much evolution 
of the cosmic dust attenuation.

Thus, we can  conclude that the bulk of the variation of 
$\rho_{\rm TIR}/\rho_{\rm FUV}$ with $z$ can be explained by 
{\sl the increase of dust attenuation with the bolometric luminosity of 
galaxies and the brightening of the galaxies at high $z$} and that we do not need a 
global increase of dust attenuation in galaxies with a fixed luminosity. 
Moreover, the evolution of luminosity functions in TIR and FUV does not exclude a 
slight diminution of $L_{\rm TIR}/L_{\rm FUV}$ and hence of dust attenuation 
in individual galaxies from $z=0$ to $z=0.7$ ($ \Delta (A_{\rm FUV}) \simeq 
0.5$~mag).

 Note however that the much lower dust attenuation found in luminous UV selected galaxies at z= 1-2 as compared to z=0 risks to be at odd with the evolution of $\rho_{\rm TIR}/\rho_{\rm FUV}$ with z as discussed above since the variation of $\rho_{\rm TIR}/\rho_{\rm FUV}$ does not seem to leave  room for a strong decrease of  dust attenuation in galaxies forming the bulk of the FUV and TIR luminosity densities. Nevertheless a substancial fraction of the galaxy samples of \citet{reddy} and \citet{burgarella} are not detected in infrared and  these non-detections must be accounted for a complete statistical analysis. 
A comparison at the same redshift of well controlled FUV and TIR (rest-frame) selected samples with a high detection rates at both wavelength (FUV and TIR) and for both samples will help to solve these issues. Such an analysis  is in progress (Takeuchi  et al. in preparation).\\

\section{Conclusions}
   We have measured the FUV ($\sim$ 1500 $\AA$) rest-frame emission of a sample of 190 LIRGs at z$\sim$0.7, 83$\%$ of these galaxies are detected in FUV. The ratio of the total infrared luminosity to the FUV one, $L_{\rm TIR}/L_{\rm FUV}$,  is compared to that found for local galaxies of similar luminosity and also selected in far-infrared. The two samples at z=0 and z=0.7 are found to  be drawn from different parent populations with a broader distribution of $L_{\rm TIR}/L_{\rm FUV}$ at z=0.7 and the presence of LIRGs with a moderate dust attenuation ($A({\rm FUV}<3$ mag) as traced by $L_{\rm TIR}/L_{\rm FUV}$. A slight difference is found for  $<L_{\rm TIR}/L_{\rm FUV}>$ between the two samples (0.2 dex) implying a decrease of the mean dust attenuation of  $\sim 0.5$ mag from z=0 to z=0.7. A lower dust attenuation in LIRGs at medium z might be related to  the  decrease of their  metallicity. One may also invoke the increased fraction of spiral-like objects in the LIRG population from z=0 to $z>0.5$ found by  several authors. If we assume that disturbed objects are more affected by dust attenuation than undisturbed ones, dust attenuation is expected to decrease when z increases.\\
Nevertheless, the amplitude of the variation of the dust attenuation  obtained for these infrared selected, bright galaxies  is  much lower than that reported for galaxies of similar luminosities at z=1-2, selected in UV-optical and detected in thermal infrared.\\
The intrinsic brightening of galaxies when  redshift increases  together with the well established variation of  dust attenuation with the luminosity of  galaxies  explain the increase of $\rho_{\rm TIR}/\rho_{\rm FUV}$ from z=0 to z=0.7. A slight diminution of  dust attenuation  in bright galaxies as observed here for the LIRG population remains consistent with the observed evolution of $\rho_{\rm TIR}/\rho_{\rm FUV}$. \\

\begin{acknowledgements}
V.B and D.B.  gratefully acknowledge ''Programme National Galaxie'' and ''Programme National Cosmologie'' support for GALEX/SPITZER-MIPS science analysis. 
TTT has been supported by  
The 21st Century Center-of-Excellence Program
``Exploring New Science by Bridging Particle-Matter Hierarchy'', Tohoku
University. Support for ELF's work was provided by NASA through the Spitzer Space Telescope Fellowship Program.

\end{acknowledgements}

\end{document}